\documentclass[a4paper]{article}

\usepackage{INTERSPEECH2020}
\usepackage{color,xcolor}
\usepackage{tipa}
\usepackage{multirow}
\usepackage{float}
\usepackage{flushend}
\usepackage[utf8]{inputenc}

\title{Why Did the x-Vector System Miss a Target Speaker?\\Impact of Acoustic Mismatch Upon Target Score on VoxCeleb Data}

\name{Rosa Gonz\'alez Hautamäki and Tomi Kinnunen}
\address{
  Computational Speech Group, School of Computing, University of Eastern Finland, Finland
\email{\{rgonza,tkinnu\}@cs.uef.fi}
}
\begin{document}

\maketitle
\begin{abstract}
Modern automatic speaker verification (ASV) relies heavily on machine learning implemented through deep neural networks. It can be difficult to interpret the output of these black boxes. In line with interpretative machine learning, we model the dependency of 
ASV detection score upon acoustic mismatch of the enrollment and test utterances. We aim to identify mismatch factors that explain target speaker misses (false rejections). We use distance in the first- and second-order statistics of selected acoustic features as the predictors in a linear mixed effects model, while a standard Kaldi x-vector system forms our ASV black-box. Our results on the VoxCeleb data reveal the most prominent mismatch factor to be in F0 mean, followed by  mismatches associated with formant frequencies. Our findings indicate that x-vector systems lack robustness to intra-speaker variations.
\end{abstract}

\section{Introduction}

\emph{Automatic speaker verification} (ASV) \cite{Kinnunen2010-overview} systems take a pair of utterances (enrolment and test) to predict if the speakers in them are same or different. When the former is actually true, such pairwise comparison is known as a \emph{target} trial, otherwise as a \emph{nontarget} trial. The prediction can be hard binary decision or a real-valued speaker similarity score. Current state-of-the-art relies largely on deep neural networks, such as the \emph{x-vector} \cite{Snyder2018-xvector} architecture, to extract speaker embeddings from each utterance. Speaker similarity score is then formed by comparing the enrolment and test embeddings using a back-end classifier \cite{PrinceE07}.

ASV systems are typically optimized to make accurate predictions for given data, \emph{on average}; not all the speakers or trials are necessarily equally treated. ASV systems are typically required to operate in an \emph{open-world} setting where the number of target speakers (classes) is allowed to increase dynamically. Additionally, ASV is used across varied operating conditions including unseen microphones, environments, and speaking styles. Thus, despite the effort that one spends on optimization, ASV systems are bound to face the unknown. Moreover, reliance on machine learning may yield decisions that humans have difficulty to interpret. The importance of explaining the decisions of machine learning systems is acknowledged and ASV is no exception. Forensic voice comparison is a canonical example of a high-stakes application where importance of explainable decisions is evident. Nonetheless, explaining the decisions of ASV systems is important for researchers, too, as it may reveal system loopholes.

We model the dependency of ASV score upon acoustic mismatch in enrollment and test data. The ASV system is treated as a black-box given \emph{as-is}: we may run it on new speech data to obtain speaker similarity scores, but otherwise we cannot optimize or interact with it. The acoustic features, however, are selected by us based on hypotheses on the type of variation expected in given data. Our work is reminiscent of \emph{score calibration} \cite{Mandasari2013-calibration} where ASV score is adjusted with the aid of external quality signals as side information. Nonetheless, besides using different methodology \cite{GonzalezHautamaki2019-JASAdisguise}, our perspective is on \textbf{explanatory analysis of a ASV system on a given evaluation corpus}, rather than on improving predictive performance. Other related research includes probing information in speaker embeddings \cite{Wang2017,Ra2019-probing}. Different from these studies that are either specific to a given type of speaker embedding or require training new classifiers in the embedding space, we model the detection score in terms of explanatory variables. The latter consists of acoustic-phonetic measures available in a public-domain toolkit \cite{openSMILE2013}.


We focus on modeling target trials. The ideal ASV score for a same-speaker (target) trial is as large number as possible --- optimally, plus infinity. Acoustic mismatch between enrollment and test data may lower the ASV score and consequently lead to falsely rejected (missed) target speaker. In an access control context, miss implies user inconvenience and in a forensic context it implies falsely declaring that the perpetrator is not present in a given trace sample. Using an up-to-date x-vector system and the large-scale VoxCeleb dataset \cite{Nagrani19} that consists of `found data' quality celebrity recordings, we aim to identify what types of acoustic mismatches are likely to contribute to increased target speaker misses.

We extend upon our recent work \cite{GonzalezHautamaki2019-JASAdisguise} in terms of speech database size, qualities, and the selected acoustic features. In \cite{GonzalezHautamaki2019-JASAdisguise} we used a self-collected (now publicly available) AVOID corpus of 60 Finnish speakers. The speakers were asked to purposefully modify their voices to sound like \emph{old} and \emph{child} speakers, so as to purposefully reinforce large variation between enrollment and test data. Indeed, the standard Kaldi x-vector system was shown to severely degrade. For instance, \emph{equal error rate} (EER) of male speakers increased from $\sim1.6\%$ (modal-modal) to $\sim25\%$ (modal-intended child). The degradation was associated/explained by differences in F0 and formants. Nonetheless, one may argue that in contemporary communication context, we do not attempt to disguise our identity or perform caricaturic voice acting. Nonetheless, the authors have observed substantial variation in speaking styles and background audio qualities in VoxCeleb data through informal listening. It is therefore plausible that target speakers may get easily missed on VoxCeleb data, too. Thus, another aim of our work is to address generalizability of our earlier findings  \cite{GonzalezHautamaki2019-JASAdisguise} (for acted voice data) to contemporary speech present in the VoxCeleb dataset.


\section{Analysis methodology}

We provide a brief summary of the interpretative model presented in \cite{GonzalezHautamaki2019-JASAdisguise}. An important aspect of ASV systems reliability is to understand the factors that affect its accuracy.  Can the score provided by ASV systems be explained by changes in acoustic measures of compared speech segments? To address this question, we model our data using a statistical regression technique specially design for repeated measures known as \emph{linear mixed effect model} (LME) \cite{lme4}. In general, regression models seek to relate a \emph{dependent variable} to a set of \emph{predictors} or \emph{independent variables}. 

\subsection{Dependent and Predictor Variables}

Let $\mathcal{U}=(\mathcal{U}_\text{e},\mathcal{U}_\text{t})$ denote a pair of enrollment and test utterances. An ASV system produces a \emph{log-likelihood ratio} (LLR) score (\emph{dependent variable}, $y$) between the two utterances as,
    \begin{equation}\label{eq:asv-llr}
        y = \log \frac{p(\mathcal{U}|H_0,\vec{\theta}_\text{asv})}{p(\mathcal{U}|H_1,\vec{\theta}_\text{asv})},
    \end{equation}
where $H_0$ and $H_1$ represent the target (same-speaker) and nontarget (different-speaker) hypotheses, respectively, and $\vec{\theta}_\text{asv}$ encapsulates all the ASV parameters. In our case, \eqref{eq:asv-llr} represents LLR score from a \emph{probabilistic linear discriminant analysis} (PLDA) back-end classifier \cite{PrinceE07}, while the two utterances are represented using their x-vector \cite{Snyder2018-xvector} speaker embeddings. 
The higher the value of $y$, the more confident the ASV system is that the speakers in the two utterances are the same. 

While $y$ serves as the response variable, our predictor variables, $x$, are formed by \emph{acoustic distances} of the form $x = |\varphi(f(\mathcal{U}_\text{e}))-\varphi(f(\mathcal{U}_\text{t}))|$. Here $f(\cdot)$ is a short-term (frame-level) feature extractor that converts a speech utterance into a sequence of scalar features, and $\varphi(\cdot)$ is a fixed summary statistics operator. By including different features and summary operators, we come up with a vector of $D$ acoustical predictors, $\vec{x}=(x_1,\dots,x_D)$ for any utterance pair $(\mathcal{U}_\text{e},\mathcal{U}_\text{t})$. In this work, $\varphi \in \{\texttt{mean},\texttt{std}\}$ consists of mean and standard deviation while the features include various standard speech features (see Table \ref{tab:feature_list}).

\subsection{Mixed effects model}

In LME models \cite{lme4}, predictors that are common to all observations are known as \emph{fixed effects}. They are represented by means of contrast. In our model, these are the acoustic distances for each single target trial. Factors that are considered as a sample of a population, in turn, are known as \emph{random effects}. The random effects in our model are the speakers. The model reflects variations associated with the speakers, as a variable with zero mean and unknown variance.

To be more specific, our model is defined as:
\begin{equation}\label{eq:randomintercept}
y_{ij} = \boldsymbol{\beta}^t \vec{x}_{ij} + b_i + \varepsilon_{ij},
\end{equation}
where $y_{ij}$ is the LLR score for the $j$th trial of target speaker $i$, $\boldsymbol{\beta}^t \vec{x}_{ij}$ is the fixed effect part (acoustic distances and their weights), $b_i$ is the per-speaker \textit{random effect} and $\varepsilon_{ij}$ is the residual. The assumption for a random speaker effect and the residual error is that they are independent of each other and follow a normal distribution: $b_i \sim \mathcal{N}(0, {\sigma_b}^2)$ and $\varepsilon_{ij} \sim \mathcal{N}(0, {\sigma}^2).$

\section{Experimental setup}
\subsection{VoxCeleb corpus}
\emph{VoxCeleb} is a publicly available large-scale dataset of speech extracted from celebrities' YouTube videos~\cite{Nagrani19}. {VoxCeleb1} contains over 100,000 utterances from 1251 celebrities with 55\% male speakers. \emph{VoxCeleb2}, in turn, contains over 6000 celebrities (61\% male). VoxCeleb2 is mainly used as a training set for ASV systems evaluations. The audio material can be considered as real-world \emph{found data} including a variety of background noises, audio quality from different processing and recording devices, and speech style variations. It mostly consists of interviews in radio and TV programs, theaters, and red carpet. In the present study, we analyze the speaker variation of the entire {VoxCeleb1}'s dataset, with speech from all the 1251 speakers,  561 female and 690 male, comprising 121,350 and 168,571 same speaker trials respectively. In contrast to our previous study where speakers were asked to disguise their voices \cite{GonzalezHautamaki2019-JASAdisguise}, the speech variations in the VoxCeleb dataset correspond to the circumstances in which they are performed --- whether a live-show interview with an audience, a radio or TV program in a formal or informal atmosphere. 

\subsection{ASV system}
X-vector embedding \cite{Snyder2018-xvector} is based on speaker-discriminative training of a deep neural network model with a long temporal context. The x-vector system uses 30 mel-frequency cepstral coefficients (MFCCs) as input features, extracted from 25 ms frames, mean-normalized over a sliding window of three seconds. Non-speech frames are discarded with an energy-based speech activity detection. For speaker similarity scoring, \emph{probabilistic linear discriminant analysis} (PLDA) is used as back-end~\cite{PLDA_Original}.
In practice, we use the pre-trained x-vector recipe in Kaldi  \cite{Povey_ASRU2011} trained on augmented VoxCeleb2 dataset \cite{Snyder2018-xvector}. Scoring this system on VoxCeleb1  (VoxCeleb1-E trial list) results in equal error rate (EER) of 2.54\%.

\subsection{Acoustic features}
\begin{table}[!hb]
\caption{The mixed effect model uses a total of 23 predictor features, formed from the following combinations of features and their long-term statistical summary measures.}
\centering
\begin{tabular}{|l|p{3.5 cm}|c|}
\hline
\multicolumn{3}{|c|}{Acoustic features, $f$} \\ \hline\hline
F0&\multicolumn{1}{|l}{Fundamental frequency } & \multicolumn{1}{l|}{F0} \\  \hline
\multicolumn{1}{|l|}{\multirow{6}{*}{VQ}} &\multicolumn{1}{l}{Loudness  } & \multicolumn{1}{l|}{ } \\  
&\multicolumn{1}{|l}{Jitter  } & \multicolumn{1}{l|}{ } \\  
&\multicolumn{1}{|l}{Shimmer  } & \multicolumn{1}{l|}{} \\ 
&\multicolumn{1}{|l}{log Harmonic-to-noise-Ratio} & \multicolumn{1}{l|}{HNR } \\  
&\multicolumn{1}{|l}{Spectral tilt } & \multicolumn{1}{l|}{H1 -- H2} \\
&\multicolumn{1}{|l}{  } & \multicolumn{1}{l|}{ H1 -- A3 } \\ 
\hline
\multicolumn{1}{|l|}{\multirow{2}{*}{Formant }}&\multicolumn{1}{|l}{Formant frequencies, } & \multicolumn{1}{l|}{F1 to F4}\\ 
&  \multicolumn{1}{|l}{formant bandwidths, } & \multicolumn{1}{l|}{B1 to B4} \\
&  \multicolumn{1}{|l}{formant amplitudes  } & \multicolumn{1}{l|}{A1 to A4 } \\
  \hline
Spectral f.&\multicolumn{1}{|l}{Spectral flux  } & \multicolumn{1}{l|}{ } \\  \hline
\multicolumn{1}{|l|}{\multirow{3}{*}{Temporal}}&\multicolumn{1}{|l}{Voiced segments per second  } & \multicolumn{1}{l|}{ }\\
&\multicolumn{1}{|l}{Voiced segments length} & \multicolumn{1}{l|}{ } \\
&\multicolumn{1}{|l}{Unvoiced segments length} & \multicolumn{1}{l|}{ }\\  \hline

\end{tabular}
\label{tab:feature_list}
\end{table}

The selected acoustic features presented in Table \ref{tab:feature_list} were extracted automatically using the openSMILE toolkit \cite{openSMILE2013}, which implements feature extraction at the frame level and provides summarization through statistical functionals at the utterance level. It has been used to serve applications such as emotion recognition, speaker trait analysis, and speaker recognition. Automatic extraction of features allows analysis of large datasets (such as VoxCeleb) for which phonetic annotations are not available. 
Even if feature extraction is performed without supervision (such as hand-made post-corrections), we expect a reliable summary of the prominent acoustic variations presented in the VoxCeleb data. The selected feature extraction parameters are based on earlier work in the analysis of voice production changes related to affective states. This feature set is known as the \emph{extended Geneva Minimalistic Acoustic Parameter Set} (eGeMAPS) \cite{egemaps_2016}.    
The selected set of 23 acoustic features can be grouped as follows: F0, voice quality (VQ), formant, spectral flux and temporal.

\subsection{Modeling the effect of acoustic variations}
\begin{figure}[!t]
 \centering
\begin{minipage}[b]{1.0\linewidth}
   \centerline{\includegraphics[width=0.7\linewidth]{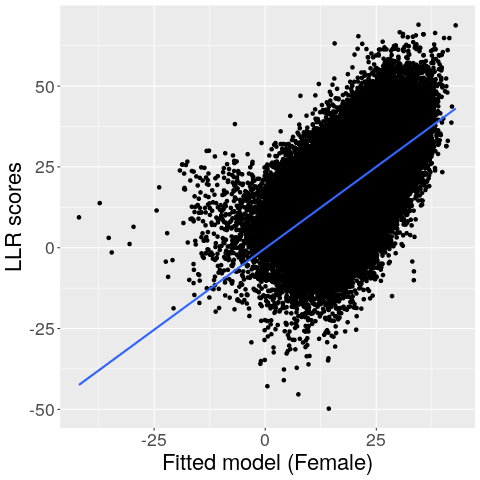}}
  \vspace{-0.1cm}
  \centerline{(a)}\medskip
\end{minipage}
\begin{minipage}[b]{1.0\linewidth}
  \centering
  \centerline{\includegraphics[width=0.7\linewidth]{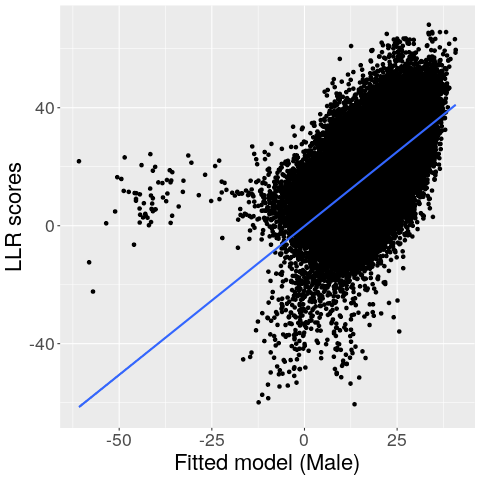}}
  \vspace{-0.1cm}
  \centerline{(b)}\medskip
\end{minipage}
\caption{Correlation for fitted model values and LLR scores (x-vector) with Pearson correlation $r$=0.60 for female and $r=0.58$ for male speakers' trials }
\label{fig:fitted_models}
\vspace{-0.3cm}
\end{figure}
We investigate the effect that acoustic feature variation at the speaker level have on LLR score of target trials. Each trial was represented by the absolute difference of the mean and standard deviation of the features in Table~\ref{tab:feature_list}. The two summary statistics (mean and standard deviation) for all the segments in the trial list were further standardized prior to the distance (absolute difference) computation. The original acoustic features  have varied ranges and this normalization ensures that none of them dominate the distance computation.

For the LME model, the acoustic feature distances were used as the fixed effects. They were used both individually, and as groups of features. In the latter case, we simply sum up the distance values within a given feature group (example: all features within the voice quality group) to form a new predictor.

As random effects, we defined intercepts for each speaker.
In this exploratory model, we seek to identify the feature variation that better explains the LLR score per trial and formulate speaker level interpretation of this relation.  We first verified that our dependent variable, the LLR score, is approximately normally distributed, which is an assumption in our model. Visual inspection of density and quantile-quantile plots showed that even without a perfect normality, the assumption was met reasonably well for our model. We use the \emph{lme4 package} \cite{lme4} to fit the linear mixed effects model, using Wald's F-test to obtained the significance test.

\begin{table}[!t]
\caption{Parameter of the mixed effects model of x-vector ASV system's scores and acoustic feature groups variations with speakers as random effect. $r$ used for feature group ranking.}
\begin{tabular}{|llrcc|}
\hline
\multicolumn{5}{|c|}{ Male speakers }\\ 
\multicolumn{5}{|l|}{Fixed effects:}  \\ \hline
 & \multicolumn{1}{r}{Estimate} &  Std. error & \multicolumn{1}{c|}{ $t$-value}& $r$ \\ \hline
$\beta_0$: Intercept & \multicolumn{1}{r}{28.36} & 0.19 & 149.3 & \multicolumn{1}{|l|}{} \\ 
$\beta_1$: F0  & \multicolumn{1}{r}{$-1.02$} & 0.02 & $-47.81$& \multicolumn{1}{|l|}{0.54}\\ 
$\beta_2$: VQ& \multicolumn{1}{r}{$-0.36$} & 0.006 &  $-56.72$ & \multicolumn{1}{|l|}{0.54}\\ 
$\beta_3$: Formant 1& \multicolumn{1}{r}{$-0.20$} & 0.01 & $-15.41$& \multicolumn{1}{|l|}{0.52}   \\ 
$\beta_4$: Formant 2& \multicolumn{1}{r}{$-0.15$} & 0.01 & $-10.04$ & \multicolumn{1}{|l|}{0.51} \\
$\beta_5$: Formant 3& \multicolumn{1}{r}{$-0.16$} & 0.01 & $-11.11$ & \multicolumn{1}{|l|}{0.51} \\
$\beta_6$: Formant 4& \multicolumn{1}{r}{$-0.31$} & 0.01 & $-30.60$ & \multicolumn{1}{|l|}{0.50} \\
$\beta_7$: Temporal& \multicolumn{1}{r}{$-0.01$} &0.009  & $-1.56$ & \multicolumn{1}{|l|}{0.48}  \\
$\beta_8$: Spectral flux& \multicolumn{1}{r}{$-0.29$} & 0.007  & $-38.43$ &  \multicolumn{1}{|l|}{0.47} \\\hline
\multicolumn{5}{|l|}{Random effects:} \\ 
\multicolumn{1}{|c}{}&  \multicolumn{1}{c}{Variance}& & & \multicolumn{1}{l|}{}\\ \hline
 Speaker: $\sigma_b^2$  &  \multicolumn{1}{r}{$4.67^2$} & & &\multicolumn{1}{l|}{} \\ 
 Residual:$\sigma^2$  &  \multicolumn{1}{r}{$9.01^2$} &  & &\multicolumn{1}{l|}{}\\ \hline
\multicolumn{4}{l}{} \\ \hline
\multicolumn{5}{|c|}{ Female speakers }\\ 
\multicolumn{5}{||l|}{Fixed effects:}  \\ \hline
 & \multicolumn{1}{r}{Estimate} &  Std. error & \multicolumn{1}{c|}{ $t$-value} & \multicolumn{1}{c|}{$r$} \\ \hline
$\beta_0$: Intercept & \multicolumn{1}{r}{32.60} & 0.21 & 155.30 & \multicolumn{1}{|l|}{} \\ 
$\beta_1$: F0  & \multicolumn{1}{r}{$-1.01$} & 0.03 & $-38.03$ &\multicolumn{1}{|l|}{0.52}\\ 
$\beta_2$: Formant 3 & \multicolumn{1}{r}{$-0.37$} & 0.02 & $-22.31$ & \multicolumn{1}{|l|}{0.52}  \\ 
$\beta_3$: VQ & \multicolumn{1}{r}{$-0.25$} & 0.008 & $-32.56$ &\multicolumn{1}{|l|}{0.52}  \\ 
$\beta_4$: Formant 2 & \multicolumn{1}{r}{$-0.41$} & 0.02 & $-23.96$ &  \multicolumn{1}{|l|}{0.52} \\ 
$\beta_5$: Formant 1 & \multicolumn{1}{r}{$-0.29$} & 0.01 & $-19.55$ & \multicolumn{1}{|l|}{0.51} \\
$\beta_6$: Formant 4 & \multicolumn{1}{r}{$-0.32$} & 0.01 & $-26.43$ &\multicolumn{1}{|l|}{0.51}  \\
$\beta_8$: Spectral flux & \multicolumn{1}{r}{$-0.29$} & 0.009 & $-32.51$ & \multicolumn{1}{|l|}{0.47} \\
$\beta_7$: Temporal & \multicolumn{1}{r}{$-0.13$} & 0.01 & $-12.36$ &\multicolumn{1}{|l|}{0.47}  \\
\hline
\multicolumn{5}{|l|}{Random effects:} \\ 
\multicolumn{1}{|c}{}&  \multicolumn{1}{c}{Variance}& & &\multicolumn{1}{l|}{}\\ \hline
 Speaker: $\sigma_b^2$  &  \multicolumn{1}{r}{$4.5^2$} & && \multicolumn{1}{l|}{} \\ 
 Residual:$\sigma^2$  &  \multicolumn{1}{r}{$9.3^2$} &  & &\multicolumn{1}{l|}{}\\ \hline
\end{tabular}
\label{model:speaker_effect}
\vspace{-0.3cm}
\end{table}

\begin{table*}[!h]
\caption{Correlation of fitted models from individual differences from Formant features (mean and standard deviation (SD)  and LLR scores}
\begin{tabular}{|l|rr|rr|rr|rr|rr|rr|rr|rr|}
\cline{2-17}
\multicolumn{1}{c|}{} & \multicolumn{8}{c|}{Male}  & \multicolumn{8}{c|}{Female} \\ \cline{2-17}
\multicolumn{1}{c|}{}  & \multicolumn{2}{c|}{Formant 1}  & \multicolumn{2}{c|}{Formant 2} & \multicolumn{2}{c|}{Formant 3}  & \multicolumn{2}{c|}{Formant 4}& \multicolumn{2}{c|}{Formant 1}  & \multicolumn{2}{c|}{Formant 2}& \multicolumn{2}{c|}{Formant 3}  & \multicolumn{2}{c|}{Formant 4}\\ \hline
\multicolumn{1}{|c|}{\multirow{1}{*}{$\mu$}} & F1 & 0.490 & A2 & 0.480 & A3 & 0.481 & A4 & 0.482 & F1 & 0.472 & A2 & 0.474 & A3 & 0.475 & B4 & 0.478 \\ 
 & \multicolumn{ 1}{r}{A1} & 0.477 & F2 & 0.474 & F3 & 0.463 & B4 & 0.471 & A1 & 0.470 & F2 & 0.459 & F3 & 0.462 & A4 & 0.476 \\ 
 & \multicolumn{ 1}{r}{B1} & 0.465 & B2 & 0.469 & B3 & 0.466 & F4 & 0.457 & B1 & 0.460 & B2 & 0.455 & B3 & 0.454 & F4 & 0.463 \\ \hline\hline
\multicolumn{1}{|c|}{$\sigma$} & \multicolumn{ 1}{r}{F1} & 0.474 & A2 & 0.462 & A3 & 0.466 & A4 & 0.468 & F1 & 0.455 & F2 & 0.466 & F3 & 0.463 & F4 & 0.474 \\ 
 & \multicolumn{ 1}{r}{B1} & 0.462 & F2 & 0.460 & F3 & 0.464 & F4 & 0.468 & A1 & 0.450 & A2 & 0.455 & A3 & 0.459 & B4 & 0.471 \\ 
 & \multicolumn{ 1}{r}{A1} & 0.459 & B2 & 0.454 & B3 & 0.451 & B4 & 0.459 & B1 & 0.443 & B2 & 0.442 & B3 & 0.450 & A4 & 0.462 \\ \hline
\end{tabular}
\label{tab:cors_formants}
\vspace{-0.3cm}
\end{table*}

\subsection{Metrics}
To evaluate the feature distances in terms of their added information to the model, we compared the correlation of the fitted model of individual feature distances with the LLR scores using Pearson correlation.  The feature distances were ranked based on how their inclusion in the model increased the correlation of fitted model and LLR score.  Also different models with groups of feature distances were compared using a standard likelihood test \emph{ANOVA}.  The \emph{Akaike information criterion} (AIC)~\cite{Akaike1974} value was used to compare the models and identified the model with better fit.  The AIC value decreases with better models.

\begin{table}[H]
\caption{Correlation of fitted models from individual voice quality (VQ) feature (mean and standard deviation) and LLR scores}
\centering
\begin{tabular}{|l|lr|lr|}
\hline
 & \multicolumn{2}{c|}{Male} &  \multicolumn{2}{c|}{Female} \\ \hline
\multicolumn{1}{|c|}{\multirow{6}{*}{$\mu$}} & \multicolumn{ 1}{l}{HNR}&\multicolumn{ 1}{l|}{0.498}&\multicolumn{ 1}{l}{H1-A3}& 0.493 \\
& \multicolumn{ 1}{l}{H1-A3} & 0.497 & HNR & 0.478 \\
 & \multicolumn{ 1}{l}{Loudness} & 0.487 & Loudness & 0.474 \\ 
 & \multicolumn{ 1}{l}{H1-H2} & 0.470 & H1-H2 & 0.456 \\ 
 & \multicolumn{ 1}{l}{Shimmer} & 0.457 & Shimmer & 0.450 \\
 & \multicolumn{ 1}{l}{Jitter} & 0.454 & Jitter & 0.444 \\ \hline \hline
\multicolumn{1}{|c|}{\multirow{6}{*}{$\sigma$}} 
 & \multicolumn{ 1}{l}{Loudness} & 0.473 & Loudness & 0.464 \\ 
 & \multicolumn{ 1}{l}{HNR} & 0.471 & HNR & 0.458 \\ 
 & \multicolumn{ 1}{l}{Shimmer} & 0.460 & Shimmer & 0.456 \\ 
 & \multicolumn{ 1}{l}{H1-H2} & 0.453 & H1-H2 & 0.440 \\ 
 & \multicolumn{ 1}{l}{H1-A3} & 0.452 & H1-A3 & 0.440 \\
 & \multicolumn{ 1}{l}{Jitter} & 0.448 & Jitter & 0.437 \\ \hline
\end{tabular}
\label{tab:cors_vq}
\vspace{-0.3cm}
\end{table}

\section{Results}
We analyze the change of acoustical features to explain the LLR score associated with the target trial's enrollment and test utterances of VoxCeleb1 data separated by gender. 
We fitted linear mixed effect models with the sum of feature distances corresponding to the feature group variation.  The eight feature group distances models were fitted with speakers as the random effects. We compared the feature group models using the Pearson correlation between fitted  values  of the model and the LLR scores, a higher correlation coefficient ($r$) indicated the order in which the feature group were added the final model.
Table \ref{model:speaker_effect} presents the regression coefficients for the final models for female and male speakers separately. The $r$ coefficient was used for the ranking of the feature group in the model.

F0 is the feature group distance that contributes first to our explanatory model. It is worth mention that this feature group consist only of two measures, the F0 mean and standard deviation distances in semitone scale. While other feature groups consist of six to twelve feature distances with exception of spectral flux that also includes two distance measurements. 

Figures \ref{fig:fitted_models} shows the correlation between the fitted model values and the x-vector's LLR scores.  Visual inspection of residual plots did not reveal obvious deviations from homoscedasticity or normality. AIC and p-values were obtained by maximum likelihood ratio tests. Both gender models have a similar correlation coefficient, $r$ of 0.6 for female and 0.58 for male speakers.  The lower correlation coefficient is expected considering the variability not dependent on the speaker effect is high with residual error variation of $9.3^2$ for females and $9.01^2$ for males. The variation corresponding to the speaker effect is similar for female and male speakers, $4.5^2$ and $4.67$ respectively. Since all the trials' LLR scores were used in this exploratory model it is expected that some observations could be consider as "outliers" enabling the identification of a group of speakers' trials to further analyze. 

\subsection{Ranking of features in terms of their explanatory power}

The feature ranking was based on the highest Pearson correlation between the model fitted with the feature groups and the LLR scores as shown in column $r$ in Table \ref{model:speaker_effect}. To analyze the importance of variations for independent feature's distances, models were fitted with each feature and the correlation to LLR scores was use to compare them. As mentioned in the previous section, F0 distance was the individual most important feature in the exploratory models. Then we analyzed the features in the voice quality and formant groups.  Table \ref{tab:cors_formants} shows the ranking of formant features (frequency, bandwidth and amplitude) distances. For both genders, amplitudes and frequencies provide more information to the model in their respective formant group for mean and standard deviation of the feature distance. 

Similar analysis was carried out for the voice quality features.  Harmonic-to-noise-ratio and harmonic variation H1-A3 provided more information to the model as shown in the correlation of the fitted models and the LLR score presented in Table \ref{tab:cors_vq}.  The ranking was nearly consistent for both genders, being shimmer and jitter the lowest ranked feature distances in this feature group.



\section{Conclusion}

Why does a given automatic speaker verification system miss (reject) a target speaker? Ideally this should not happen in the first place, but when it does, it is useful to analyze the reasons. This may  suggest ideas for future improvements of the recognition technology itself, inform users of the limitations of a given recognizer, or suggest ways of composing new evaluation corpora based on \emph{found data}. No automatic speaker verification system is (or will likely ever be) completely immune to mismatch across enrollment and test data.

We approached the question from the perspective of regression analysis using a linear mixed effects model. The modeled variable is the LLR score of a speaker recognition system (here, x-vector PLDA) while the predictor variables consist of enrollment-vs-test distances in the first-order (mean) and second-order (standard deviation) statistics of selected acoustic features. We extended our previous work \cite{GonzalezHautamaki2019-JASAdisguise} in terms of the database and the acoustic features. 

Overall, the acoustic variation impacts strongly the score of the ASV system. We found correlations up to $\approx 0.6$ of the fitted model and the LLR score. Interestingly, our analysis confirms an important  finding noted in \cite{GonzalezHautamaki2019-JASAdisguise} for a completely different corpus (but the same, Kaldi x-vector system): F0 mismatch plays a key role. Unsurprisingly, differences in formants and voice quality parameters contribute to degraded score, too. 

\bibliographystyle{IEEEtran}
\bibliography{mybib}

\end{document}